\documentclass[aps,amsmath,prb,floatfix,twocolumn]{revtex4}
\long\def\comment#1{}
\usepackage{graphicx}
\usepackage{amsmath}
\usepackage{amsfonts}
\usepackage{amssymb}
\usepackage{color}

\newcommand{\beq}{\begin{equation}}
\newcommand{\eneq}{\end{equation}}
\newcommand{\bea}{\begin{eqnarray}}
\newcommand{\enea}{\end{eqnarray}}

\begin{document}
\title{High critical temperature nodal superconductors as building block for time-reversal invariant topological superconductivity}

\author{F. Trani$^{1,2}$}
\email{fabio.trani@spin.cnr.it}
\author{G. Campagnano$^{2,1}$}
\author{A. Tagliacozzo$^{1,2}$}
\author{P. Lucignano$^{2,1}$}

\affiliation{$^1$ Dipartimento di Fisica "E. Pancini", Universit\`a di Napoli ``Federico II'', Monte S.Angelo, I-80126 Napoli, Italy}
\affiliation{$^2$ CNR-SPIN, Monte S.Angelo -- via Cinthia,  I-80126 Napoli, Italy}

\begin{abstract}
We study possible applications of  high critical temperature nodal  superconductors for the search for Majorana bound states in the DIII class.
We propose a microscopic analysis of the proximity effect induced by d-wave superconductors on a semiconductor wire with strong spin-orbit coupling.
We  characterize the induced superconductivity on the wire employing a numerical self-consistent tight-binding Bogoliubov-De-Gennes approach, and analytical considerations on the Green's funtion.  The order parameter induced on the wire, the pair correlation function and the renormalization of the Fermi points are analyzed in  detail, as well as the topological phase diagram in the case of weak  coupling. We highlight optimal Hamiltonian parameters to access the non trivial topological phase which could display time-reversal invariant Majorana doublets at the boundaries of the wire.
\end{abstract}

\date{\today}

\maketitle 
\section{Introduction}
After the theoretical proposal by Kitaev\cite{Kitaev:2001},  the race for the search for Majorana Bound States (MBSs) in solid state devices, is producing interesting theoretical and experimental results \cite{Alicea:2012,Beenakker:2013,DasSarma:2015}.
MBSs have been predicted in a wide class of low-dimensional solid state devices. 
For instance, they are expected to appear in conventional superconductors in contact with topological insulators (TIs) \cite{Fu:2008}, quasi one-dimensional systems with strong spin-orbit interactions  \cite{Lutchyn:2010,Oreg:2010,Duckheim:2011,Weng:2011},  helical magnets \cite{Kjaergaard:2011} and other materials \cite{Alicea:2010,Fu:2009,Akhmerov:2009,Tanaka:2009,Linder:2010,Lucignano:2012}. 
Also, the peculiar features of MBSs arising at the interface between a topological superconductor and an interacting one-dimensional electron liquid \cite{a1,a2}, have been recently discussed using an adapted version of the field theoretical approach of Refs\cite{b1,b2,b3}. 
Most of the quoted proposals resort to external magnetic fields or magnetic materials, in order to get rid of the unwanted Kramers degeneracy. 
The same is true for recent experimental realizations \cite{Kouwenhoven:2012,Heiblum:2012,Marcus:2013,VanHarlingen:2013}, mostly focused onto systems with explicit time-reversal symmetry breaking.
They all belong to ''class D'' according to the mathematical classification of Bogoliubov de Gennes Hamiltonians \cite{Schnyder:2008,Teo:2010,Qi:2010}.
However, the presence of external magnetic fields, that have to be finely tuned in order to satisfy the topological criterion without suppressing the proximity gap, poses limitations to the operating temperature, to the device geometries, and confines experiments to  limited range of materials.
On the other hand, a recent paper by Zhang and coworkers \cite{Zhang:2013} investigates a different class  (DIII) of time-reversal invariant (TRI) topological superconductors (TS). Their idea is to utilize proximity effect devices which combine Rashba semiconductors (RS) and superconductors with  $s_\pm$  or $d_{x^2-y^2}$ spin-singlet pairing potential that switches sign between the $\Gamma$ and $M$ points, whose boundary excitations are Majorana doublets (MD)s.
However a detailed microscopic analysis of the proximity effect on such system is still lacking, and the stability of the topological phases expected in these TRI topological superconductors (TRITS) has to be studied in detail. This is exactly the point that we address in our research.

High critical temperature superconductors (HTS)  have been proposed as a key building block to experimentally produce MBSs, since the high critical temperature may induce a robust superconductivity by proximity effect\cite{Takei2013,Lucignano:2012,referee_1}.  More importantly, d-wave superconductors can induce an extended s-wave superconductivity, and this can be the main ingredient for the production of TRI MDs \cite{Zhang:2013,Haim:2016}. On the other way around, Majorana quasiparticles on the wire may couple to the nodal fermions in d-wave superconductors, and leaking of Majorana states  in the substrate has been theoretically predicted\cite{Wong2012}.  This motivated us to deepen our understanding of the proximity effect in semiconductor-superconductor heterostructures.

Theoretical literature on TRITS mainly focuses on (i) intrinsic superconductivity, or (ii) proximity induced superconductivity. In the first case, the topological phase diagram and the prediction of Majorana states at the boundaries are analyzed upon changing the experimental conditions (geometry, spin-orbit coupling, gate potential, disorder, symmetry of the order parameter) \cite{Oreg:2010,Zhang:2013,Wong2012,Liu2014,Brouwer:2011} and criteria for the recognition of MBSs are proposed \cite{Sticlet2012, Sedlmayr2015}. In the second case, most of papers adopt an analytical model for the self-energy to represent the proximity-induced superconductivity \cite{Tkachov2013,Chevallier2013,McMillan1968,Potter:2011,Takei2013,Haim:2016}.

In this paper we present a detailed study of the physics of semiconductor nanowires in the presence of Rashba spin-orbit coupling \cite{Bercioux:2015}, such as InAs or InSb in proximity to a cuprate high critical temperature $d_{x^2-y^2}$ superconductor (such as YBCO), employing both a microscopic self-consistent approach based onto a tight-binding Bogolubov De Gennes (TBBDG) scheme and a semi-analytical formulation, based on a path-integral scheme, allowing us to calculate explicitly the exact proximity self-energy.
A deep analysis of the correlation effects induced on the wire, the induced gap, the excitation spectrum and the projected density of states, and their implications for the formation of Majorana bound states, is reported.  

The paper is organized as follows. 
In Sec. II we introduce the model Hamiltonian  in real and momentum spaces, and we describe our numerical and analytical computational methods. In section III we discuss our results. In particular, in section IIIA, using the expression of the selfenergy calculated before, we show the main difference between the induced gap by nodal and conventional superconductors. In section IIIB we carefully analyze the peculiarities of the excitation spectrum and the pair correlation function induced on the wire by a nodal superconductor. In Sec. IIIC we discuss criteria for the formation of Majorana bound states and show a topological phase diagram in a topological weak regime. In Sec. IIID we focus on the role of spin-orbit coupling on the induced pair correlation functions. In Sec. IV we summarize the results and compare with the existing literature. 

\section{Model Hamiltonian}
\begin{figure}[htb]
\begin{center}
\includegraphics[width=0.4\textwidth]{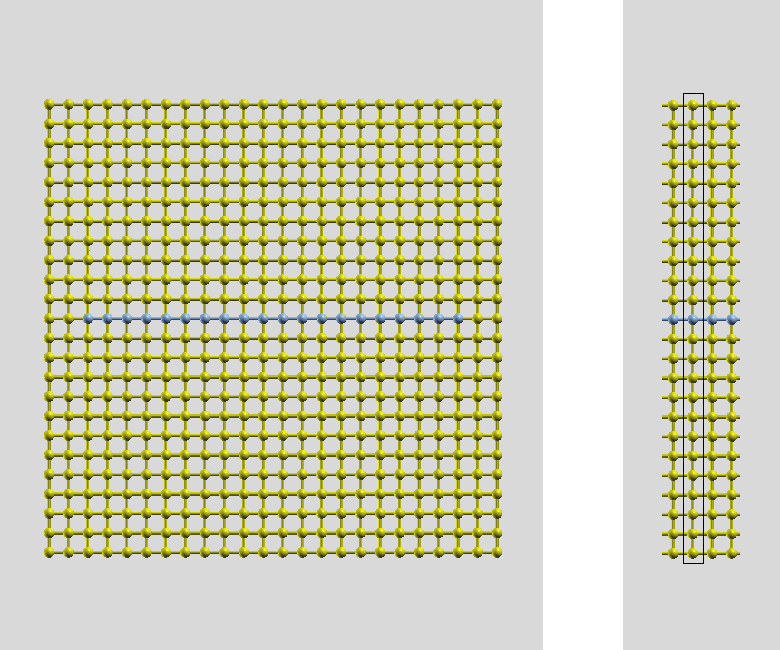} 
\end{center}
\caption{\label{fig:sys} Left panel: Top view of the system. The semiconductor wire (blue circles) lies on top of the superconductive substrate (yellow circles). 
The wire on the superconductor at angle $\theta=0$. The wire/superconductor is represented by blue/yellow spheres. Right panel:  The region inside the black rectangular box represents the unit cell.} 

\end{figure}

Our model system is formed by a semiconducting nanowire with strong spin-orbit interaction lying onto a $d_{x^2-y^2}$ superconductor.
Because of its layered structure, here we model a typical d-wave superconductor as a two-dimensional material  in the $x-y$ plane, ignoring the coupling between the underlying planes\cite{Black-Schaffer2008}. We describe the superconductor with a square lattice with periodic boundary conditions. The wire is modeled as a one-band semiconductor whose chemical potential is determined by the doping level. A  top view of the system is given in Fig. \ref{fig:sys} (left panel).

We describe the system by a tight-binding Hamiltonian composed of three terms:
\begin{equation}
H = H_s+H_w+H_{T}\:,
\end{equation}
where $H_s$ describes the $2D$ superconductor, $H_w$ describes the $1D$ wire and $H_{T}$ expresses the coupling between the two. 

\subsubsection{d-wave Superconductor}
Following Scalapino \cite{Scalapino1995} we describe our superconductor with the following mean-field Hamiltonian:
\begin{eqnarray}
&H _{s} &= -\mu_{s}  \sum _{\bf i,\sigma} c ^\dagger _{\bf i\sigma} c _{\bf i\sigma} -t_s \sum_{\left<\bf i,\bf j\right>,\sigma} c ^\dagger _{\bf i\sigma} c _{\bf j\sigma}\nonumber \\
 &&+\sum_{\bf l} \frac{\Delta_{d}(\bf l)}{2} [(c^\dagger_{\bf l+x\uparrow}c^\dagger_{\bf l\downarrow}-c^\dagger_{\bf l+x\downarrow}c^\dagger_{\bf l\uparrow})\nonumber \\ 
 &&-(c^\dagger_{\bf l+y\uparrow}c^\dagger_{\bf l\downarrow}-c^\dagger_{\bf l+y\downarrow}c^\dagger_{\bf l\uparrow})
+(c^\dagger_{\bf l-x\uparrow}c^\dagger_{\bf l\downarrow}-c^\dagger_{\bf l-x\downarrow}c^\dagger_{\bf l\uparrow})\nonumber \\ \label{eq1}
&&-(c^\dagger_{\bf l-y\uparrow}c^\dagger_{\bf l\downarrow}-c^\dagger_{\bf l-y\downarrow}c^\dagger_{\bf l\uparrow}) +\textrm{H.c.}] \:,
\end{eqnarray}
where $\bf x$, $\bf y$ are the elementary displacement. Here and in the following we set the lattice spacing $a=1$.
The first two terms describe the chemical potential $\mu_s$ and the hopping $t_s$ between nearest neighbours electron sites, while the pairing term corresponds to a  $d_{x^2-y^2}$ combination of singlets between the $l^{th}$ lattice site and its four nearest neighbours of the 2D square lattice. In what follows all the energies are calculated in units of $t_s$, except where defined otherwise. The order parameter for d-wave superconductivity is defined as
\bea 
\label{deltad}
\Delta_d (\bf l) &=& -V_d [F_{\bf l,\bf l+x}+ F_{\bf l,\bf l-x}- F_{\bf l,\bf l+y}- F_{\bf l,\bf l-y}]/4
\enea
where the singlet pairing amplitude on a bond is described by  \cite{Zhu:2000,Cuoco:2008}
\beq
\label{eqf}
F _{\bf i,\bf j} = \frac{1}{4}\langle  c_{\bf i\uparrow} c_{\bf j\downarrow} - c_{\bf i\downarrow} c_{\bf j\uparrow} + c_{\bf j\uparrow} c_{\bf i\downarrow} - c_{\bf j\downarrow} c_{\bf i\uparrow}\rangle \: .
\eneq
The order parameters are determined self-consistently from the resolution of the TBBDG equations, whereas the pair potential $V_d$ is kept at a fixed value. In Eq. (\ref{eq1}) d-wave superconductivity is considered. As a reference, we will compare our proximity superconductivity with that induced by a $s$-wave superconductor, with order parameter $\Delta_0 (\mathbf{l}) = -V_0 F_{\bf l,\bf l}$.

The bulk superconductor is investigated by Fourier-transforming Eq. (\ref{eq1}) to k-space. By defining $\mathbf{k}=(k_x,k_y)$, the Hamiltonian of the superconductor is written as
\bea
H _{s} &=&   \sum _{\mathbf{k}} \xi_s (\mathbf{k}) (c ^\dagger _{\mathbf{k}\uparrow} c _{\mathbf{k}\uparrow} + c ^\dagger _{\mathbf{-k}\downarrow} c _{\mathbf{-k}\downarrow}) +   \label{eqk} \\
&+ & \sum _{\mathbf{k}} \Delta (\mathbf{k}) [(c ^\dagger _{\mathbf{k}\uparrow} c ^\dagger _{\mathbf{-k}\downarrow} - c ^\dagger _{\mathbf{-k}\downarrow} c ^\dagger _{\mathbf{k}\uparrow}) + h.c.] \: , \nonumber
\enea
where 
$\xi_s(\mathbf{k}) = -\mu_{s} - 2  \left[\cos(k_x)+\cos(k_y)\right]$ and $\Delta (\mathbf{k})=2\Delta_d [\cos(k_x)-\cos(k_y)]$.

The pair correlation function in k-space reads
\beq
\label{eqfk}
F _\mathbf{k} =  \frac{1}{2}\langle c_{\mathbf{k}\uparrow} c_{\mathbf{-k}\downarrow}-c_{\mathbf{-k}\downarrow} c_{\mathbf{k}\uparrow}\rangle \: ,
\eneq
and the d-wave order parameter is
\beq	
\label{eqdk}
\Delta_d = -\frac{V_d}{2N_\mathbf{k}}  \sum_\mathbf{k} F_\mathbf{k} \left[\cos(k_x) - \cos(k_y)\right] \: ,
\eneq
where $N_{\mathbf{k}}$ is the number of k-points and the sum is over the first Brillouin zone. The Hamiltonian in Eq. (\ref{eqk}) is diagonalized using a standard BdG scheme\cite{Ghosal:2001}.
In the new operator basis set, the Hamiltonian in Eq. (\ref{eqk}) is represented in a matrix form, whose eigenvalues give the excitation energies of the superconductors. At each step of the self-consistent scheme, a new value of the order parameter is calculated from the eigenvectors, according to Eqs. \ref{eqfk} and \ref{eqdk}, then the Hamiltonian is updated and diagonalized again, and  a new order parameter is obtained. The scheme is repeated iteratively until the pair correlation function $F_\mathbf{k}$ reaches the self-consistency for each k-point.

\begin{figure}[!h]
\begin{center}
$\begin{array}{cc}
\includegraphics[width=0.4\textwidth]{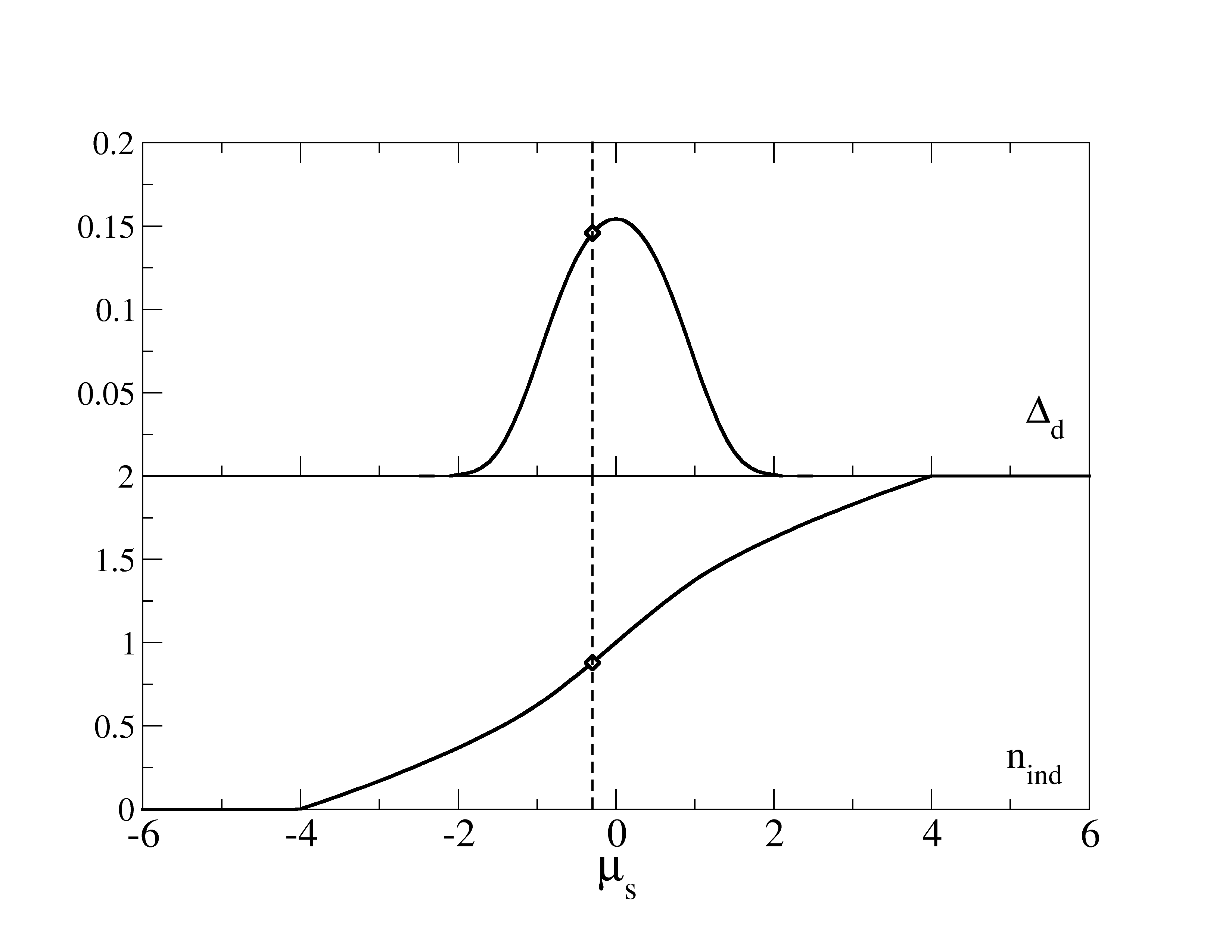} &
\end{array}$
\end{center}
\caption{\label{fig:dwave}Amplitude of superconductor d-wave order parameter (upper panel) and average electron density (lower panel) as a function of $\mu_s$, calculated using the self-consistent BdG technique. The pairing parameter used for the self-consistent calculation is:  $V_d=1.5$. }
\end{figure}

For the numerical simulations, we  fix the parameters in such a way to obtain a given reference electron density at the end of the self-consistent calculation. We choose a value $V_d = 1.5$ for the d-wave pair interaction, which is consistent with the parameter used in Ref. \onlinecite{Ghosal:2000}. In Fig. \ref{fig:dwave}, the order parameter and the electron density obtained at the end of the self-consistent scheme are reported as a function of $\mu_{s}$.
We use a value of $\mu_{s}$  maximizing the superconductor density and pairing, namely
 $\mu_s=-0.3$,  corresponding to an induced density $\langle n_{ind}\rangle=0.88$ at the end of the self-consistent calculation, in agreement with Ref. \onlinecite{Ghosal:2000}. The self-consistent  gap is $\Delta _{d}=0.15$, in agrement with the gap used in Ref. \onlinecite{Wong2012}. The convergence of all the results has been carefully verified with respect to the number of k-points.
Throughout the paper, we also discuss the trends of some physical variables, upon changing the gap, or the superconductor chemical potential. For instance, we calculate the induced gap on the wire as a function of the superconductive gap, or the topological phase diagram. In these cases, a non self-consistent (one-shot) TBBDG approach is employed at fixed values of $\Delta_d$ and $\mu_s$.

\subsubsection{Nanowire}
The semiconducting nanowire is described by the following Hamiltonian:
\begin{eqnarray}
&H _{w} &= -\mu_{w}  \sum _{i,\sigma} d ^\dagger _{i\sigma} d _{i\sigma} -t_w \sum_{\substack{<i,j>\\\sigma}} d ^\dagger _{i\sigma} d _{j\sigma} + \label{eq2} \\
&+& \imath \alpha \sum_{\substack {<i,j> \\ \sigma \sigma' } }   d ^\dagger _{i\sigma}  (  \hat\sigma_y )_{\sigma \sigma'}  d _{j\sigma'} + h.c. \: , \nonumber
\end{eqnarray}
where $\alpha$  accounts for the spin-orbit coupling of Rashba type \cite{Bercioux:2015}, and $\hat\sigma_x$, $\hat\sigma_y$ and $\hat\sigma_z$ are the spin Pauli matrices. 
We allow for a chemical potential $\mu_w$ different from that of the superconductor, and tunable by gating the heterostructure.
Superconducting correlations in the singlet channel, in our one dimensional wire, can either be of   s-wave or extended s-wave type, whose pair correlation functions, extended to the $n^{th}$ order, are defined as
\bea
F_i^{(n)} = [F_{i,i+n}+ F_{i,i-n}]/2 \:. \label{eqfcomp}
\enea
In particular, $F_i^{(0)}$ is the local $s$ component, $F_i^{(1)}$ is the extended $s$ component, $F_i^{(2)}$ is the extended next-nearest-neighbor $s$ component. The opportunity of introducing the definition in Eq. (\ref{eqfcomp}) follows by the fact that the proximity effect induces on the wire a spread of the correlation functions well beyond the local and the first s-wave extended components.
In this respect, the model of an isolated wire, where the "induced" superconductivity is put by hands, is in general very poor with respect to  our more realistic model of a wire placed on top of a superconductor, where the effective correlations induced by the proximity effect are accounted for.

\subsubsection{Superconductor + nanowire}
The coupling between the wire and the superconductor is given by a simple hopping Hamiltonian:
\begin{equation}
H_{T} = -t_{T} \sum _{\left<\mathbf{l},i\right>_z,\sigma}  c ^\dagger _{\mathbf{l}\sigma} d _{i\sigma} + h.c. \label{eq3}
\end{equation}
The hopping happens in the $z$ direction and involves  $d$ electrons, belonging to the wire, and $c$ electrons, belonging to the superconductor, and is supposed to be spin independent. 

\subsection{Computational method}
In the description of the whole system, periodic boundary conditions are assumed along the $x$ direction. The unit cell is composed by a strip, as it is reported in Fig. \ref{fig:sys} (right panel).  
The field operators are Fourier transformed along the $x$-direction. For each site $i$ of the unit cell, we calculate a superconductor pair correlation function $F_i (k)$ (allowing us to compute the superconductor gap $\Delta_{i}$), that we iteratively update until the self-consistency is obtained on each site and k-point. Once the selfconsistency on the superconductor is achieved, the pair correlation function induced on the wire, $F_{ind}(k)$, is calculated at each k-point. From $F_{ind}(k)$ the s-wave components of the pair correlation function are evaluated at all orders of neighbors.

It is convenient to introduce the Nambu notation to write the Hamiltonian. 
The field operators for the superconductor are defined as $\psi^s_{jk}=(c_{jk\uparrow},c_{jk\downarrow},c^\dag_{j-k\uparrow},c^\dag_{j-k\downarrow})$, where $j=0\ldots N-1$ labels the superconductor sites in the unit cell, and $k$ is the k vector along $x$. We introduce the field operator  $\psi^w_k=(d_{k\uparrow},d_{k\downarrow},d^\dag_{-k\uparrow},d^\dag_{-k\downarrow})$, for the wire.
The Hamiltonian is therefore written as
\begin{eqnarray}
H _{s} &=& \frac{1}{2}\sum _{jk} \psi^{s\dag} _{jk}  [\xi^s _k \hat{\tau}_z-2\Delta _{j}\cos(k)\hat\sigma_y\hat{\tau}_y]\psi^s  _{jk}  \\ 
 &+& \frac{1}{2}\sum _{jk} \left[ \psi ^{s\dag}  _{jk}  (-t _s\hat{\tau}_z + \Delta _{j}\hat\sigma_y\hat{\tau}_y )\psi^s  _{j+1k} + h.c. \right]\nonumber \\
H _{w}  &=& \frac{1}{2}\sum_k\psi ^{w\dag} _k [\xi^w_k - 2 \alpha \sin(k) \hat\sigma_y ]\hat{\tau}_z \psi^w  _k   \nonumber \\ 
H_{T} &=&\frac{1}{2} \sum_k \psi ^{w\dag}_{k}   (-t_{T}\hat{\tau}_z)\psi^s _{0k} +  h.c. \: ,\nonumber
\end{eqnarray}
where  $\xi^s_k=-\mu_s - 2 \cos(k)$, $\xi^w_k = -\mu_w - 2t_w \cos(k)$, and $\hat\tau_x$, $\hat\tau_y$ and $\hat\tau_z$ are the Nambu matrices.
The singlet pair correlation function is defined for the superconductor as  
\beq
F _{i,j}(k) = \frac{1}{4}\langle  c_{ik\uparrow} c_{j-k\downarrow} - c_{ik\downarrow} c_{j-k\uparrow} + c_{jk\uparrow} c_{i-k\downarrow} - c_{jk\downarrow} c_{i-k\uparrow}\rangle \: .
\eneq
For each superconductor site the d-wave pairing is calculated as 
\beq
\Delta _j  =-\frac{V_d}{4 N_k} \sum_k \left[ 2F _{j,j} (k) \cos(k) - F_{j,j+1} (k)  -  F_{j,j-1} (k) \right]\: .
\eneq
Starting from an initial guess of the order parameter, the  Hamiltonian of the whole system is diagonalized, the d-wave pair correlation function is calculated at each site and k-value, and the  value of the order parameter is updated in the Hamiltonian, that is diagonalized again. The procedure is iterated until the self-consistency is obtained  at each site. Once the self-consistent solutions are obtained, the  pair correlation function induced on the wire is calculated in k-space as
\beq
 F^{(w)} (k) = \frac{1}{2}\langle  d_{k\uparrow} d_{-k\downarrow} - d_{k\downarrow} d_{-k\uparrow}\rangle \: ,
\eneq
and in real space, at all order of neighbors (l=0,1,2 \ldots), it is
\beq
 F^{(w)}_l = \frac{1}{N_k}\sum _k F ^{(w)} (k) \cos(kl) \: .
\eneq
In particular, $l=0$ and $l=1$ refer to the local and extended $s$ wave contributions of the pairing induced in the wire. 

The selfconsistent results reported in the following have been performed using  $N=200$ sites in the unit cell, and a  $256$ k-points grid in the Brillouin zone. After the selfconsistent solution is found, the band structure and the pair correlation functions are calculated on a finer k-point mesh.

\subsection{Selfenergy calculation without selfconsistency}
As the hopping of Eq. \ref{eq3} is supposed to be spin independent, we can calculate the selfenergy induced in the nanowire within a restricted basis 
\begin{eqnarray}
\psi^s_{\bf k}&=&(c_{{\bf k}\uparrow},c^\dag_{{\bf -k}\downarrow}) \\
\psi^w_{k_x}&=&(d_{ k_x\uparrow},d^\dag_{ -k_x\downarrow})
\end{eqnarray}
The full partition function can be written as a fermionic path-integral \cite{NegeleOrland} as:
\begin{equation}
Z=\int \mathcal{D}[\bar{\psi}^s,\bar{\psi}^w,\psi^s,\psi^w]e^{-S[\bar{\psi}^s,\bar{\psi}^w,\psi^s,\psi^w]}
\end{equation}
with 
\begin{equation}
S=S_s+S_w+S_T \, .
\end{equation}
We have
\begin{multline}
S_s[\bar{\psi}^s,\psi^s]=\frac{1}{\beta}\sum_{\omega_n}\int_{-\pi}^{+\pi}\frac{dk_x}{2\pi} \int_{-\pi}^{+\pi}\frac{Ndk_y}{2\pi} \\ 
\times
\bar{\psi}^s_{k_x,k_y} \left[ -\imath \omega_n+\xi^s_{k_x,k_y}\hat{\tau}_z  +\Delta_{k_x,k_y}\hat{\tau}_x\right]  \psi^s_{k_x,k_y} \, ,
\end{multline}
where $\beta = 1/k_B T$ and  $\omega_n$ are the Fermionic Matsubara frequencies. Similarly
\beq
S_w[\bar{\psi}^w,\psi^w]=\frac{1}{\beta}\sum_{\omega_n}\int_{-\pi}^{+\pi}\frac{dk_x}{2\pi}  \\ 
\bar{\psi}^w_{k_x} \left[ -\imath \omega_n+\xi^w_{k_x}\hat{\tau}_z  \right] \psi^w_{k_x} \, .
\eneq
Moreover, the contribution to the action due to the tunneling Hamiltonian $H_T$ reads,
\begin{equation}
S_T=\frac{1}{\beta}\sum_{\omega_n}\int_{-\pi}^{+\pi}\!\!\frac{dk_x}{2\pi} \int_{-\pi}^{+\pi}\!\!\frac{Ndk_y}{2\pi}
\left[ \bar{\psi}^w_{k_x}  \hat{T} \psi^s_{k_x,k_y} +\bar{\psi}^s_{k_x,k_y}  \hat{T}^\dag \psi^w_{k_x}\right] \,
\end{equation}
where $\hat{T}=-t_T\hat{\tau}_z/\sqrt{N} $. In order to obtain an effective action $S^e_w$ for the wire, we integrate out 
the superconductor's degrees of freedom
to  obtain  the effective action:
\begin{equation}
\label{effact}
S^e_w=S_w + \frac{1}{\beta}\sum_{\omega_n}\int_{-\pi}^{+\pi}\frac{dk_x}{2\pi} \int_{-\pi}^{+\pi}\frac{Ndk_y}{2\pi}\hat{T}\hat{\mathcal{G}}^s_{k_x,k_y}(\imath\omega_n)\hat{T}^\dag \, ,
\end{equation}
where we have introduced the Matsubara Green's function of the superconductor 
\begin{equation}
\hat{\mathcal{G}}^s_{k_x,k_y}(\imath\omega)=\left[\imath\omega -\xi^s_{k_x,k_y}\hat{\tau}_z  - \Delta_{k_x,k_y}\hat{\tau}_x \right]^{-1}.
\end{equation}
The added term in Eq. \ref{effact}, appears as a self-energy in the wire effective Hamiltonian:
\begin{equation}\label{selfenergy}
\hat{\Sigma}(\imath\omega,k_x)=|t_T|^2\int_{-\pi}^{+\pi}\frac{dk_y}{2\pi}\left[\frac{-\imath\omega -\xi^s_{k_x,k_y}\hat{\tau}_z  +\Delta_{k_x,k_y}\hat{\tau}_x }{\omega_n^2+(\xi^s_{k_x,k_y})^2+|\Delta_{k_x,k_y}|^2} \right]\,.
\end{equation} 
This  is clearly independent of the spin-orbit coupling and the chemical potential of the wire. Starting from the self-energy, the Matsubara and retarded Green's functions of the wire can be calculated.
In the Appendix, the expressions for the analytical continuation of the Green's function and the density of the states are given. 

In the case $\Delta_0=t_s$, we can calculate exactly the self-energy of Eq.(\ref{selfenergy}) in the limit $\mu_s=0$. After the integration over $k_y$ we have
\begin{multline}
\hat{\Sigma}(\imath\omega,k_x)= \\ 
|t_T|^2 \frac{ - \imath \omega +2 t_s \cos(k_x)(\hat{\tau}_x+ \hat{\tau}_z)}{\sqrt{(8 t_s^2 \cos(k_x)^2+\omega^2)(8 t_s^2 (1+\cos(k_x)^2)+\omega^2)} }
\end{multline}

\section{Results and discussion}

\subsection{Renormalization and induced pairing}
The path-integral approach described in the previous section allows us to write down an analytic expression for the self-energy,  that accounts for an effective interaction of the superconductor on the wire. The Matsubara self-energy, reported in Eq. (\ref{selfenergy}), can be numerically computed for given sets of parameters ($\mu_s$, $\mu_w$, $t_T$, $\Delta$). In order to simplify the notation, we write the self-energy in a compact form as $\hat{\Sigma}(\imath\omega,k_x)= (-\imath\omega A - B\hat{\tau}_z+C\hat{\tau}_x)$, where A, B and C can be numerically estimated from Eq. (\ref{selfenergy}) for each value of $k_x$ and $\omega$. 
Starting from the Green's function of the wire, an expression for the induced gap $\Delta_{ind}$ can be deduced as in Ref. \onlinecite{Potter:2011}. In particular,
\beq
\label{eqgw}
 \hat{\mathcal{G}}^w (\imath\omega,k_x) = \frac{1}{ \imath\omega -\xi ^w _{k_x} \hat{\tau}_z - \hat{\Sigma}(\imath\omega,k_x) } = \frac{Z} {\imath\omega  -\tilde{\xi} ^w _{k_x} \hat{\tau}_z - \Delta_{ind}\hat{\tau}_x }
\eneq
where $Z = 1/(1+A)$ is the renormalization factor, and
\bea
\tilde{\xi}^w _{k_x} &=& Z (\xi^w - B)  \label{eq:xiwind} \\
\Delta _{ind} &=& ZC = \frac{C}{1+A}.  \nonumber
\enea
The renormalization induces a displacement of the wire's Fermi level (via the B term of the self-energy) and the formation of a superconducting gap by proximity  effect ($\Delta_{ind}$). In the case of s-wave superconductivity, we have $C=A\Delta_0$, and we  formally recover  the induced gap $\Delta_{ind} = (1-Z) \Delta_0$ reported in Ref.~\onlinecite{Potter:2011}.

\begin{figure}
\includegraphics[width=0.5\textwidth]{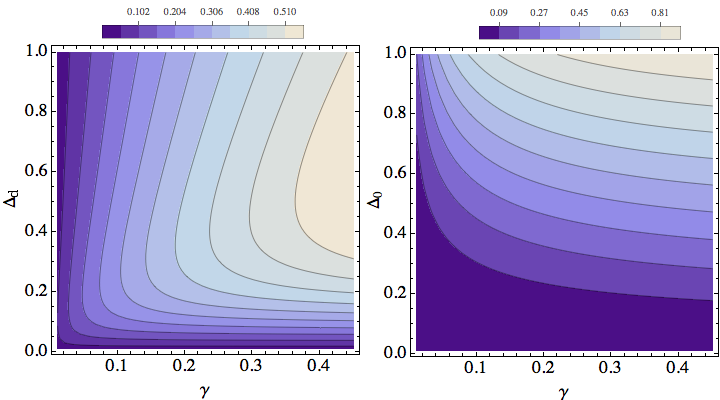}
\caption{\label{sigma}  The gap induced by proximity effect ($\Delta_{ind}$) as a function the superconductor gap ($\Delta_0$), and the coupling $\gamma$ between the superconductor and the wire for d-wave (left panel) and s-wave (right panel) order parameter in the superconductor. 
Present results were obtained using analytical expression of the self-energy, and taking into account the renormalization due to dynamical effects. 
The self-energy was evaluated at the chemical potential of the wire $\mu_w = -2 t_w \cos (k_x) | _{k_x=\pi/4}$.}
\end{figure}

In order to give a numerical estimation of the induced superconducting gap, we compute the $A$, $B$ and $C$ components of the self-energy in the static approximation ($\omega$=0).

In Fig. \ref{sigma} the induced gap is shown as a function of the superconductive gap, and of the coupling strength  $\gamma=N_B(\mu_w) t_T^2$, where $N_B(\mu_w)$ is the tunneling density of states of the superconductor at the chemical potential of the wire, in the absence of superconductivity\cite{McMillan1968}.

The left panel of Fig. \ref{sigma} reports the induced gap calculated at $k_F^w$, in the case of a d-wave order parameter, while the right one, the induced gap for a s-wave order parameter, as a comparison. 
Contour plots show that, while in the case of a conventional s-wave superconductor, the induced gap increases both as a function of $\gamma$ and as a function of $\Delta_0$, in the case of a d-wave superconductor, the induced gap increases monotonically as a function of $\gamma$ but not with $\Delta_d$. In other words, there is an optimal intermediate value of the gap amplitude in the superconductor that maximizes the induced gap in the nanowire. 
This unexpected behavior, showing a degeneration of the proximity effect in d-wave superconductors upon increasing the gap, is confirmed by the full TBBDG calculation.
To better visualize this result, we also calculate, within our analytic approach, the density of states of the wire, starting from the retarded  Green's function of the wire:
\begin{equation}
\rho^w(\omega)=-\frac{1}{2\pi}\Im m\mbox{Tr} \left(\hat{\tau}_3 \hat{\mathcal{G}}_{ret}^w\right) \,.
\end{equation}
Thus,  we need to compute the retarded self-energy that can be done by substituting $\imath\omega \rightarrow \omega+\imath 0^+$ in the Matsubara expression of 
Eq.(\ref{selfenergy}). Mathematical details are given in the Appendix. 

The same can be done also in our selfconsistent numerical tight-binding approach, after a projection of the Hamiltonian spectrum onto the wire states. Projection is, sometimes, demanding, thus we resort to our analytical densitiy of states in the following.  The results are, of course, in agreement, when selfconsistency is ignored. 
Here we plot curves obtained using the Green's function of Eq. \ref{eqgw}. 

\begin{figure}[!h]
\begin{center}
\includegraphics[width=0.49\textwidth]{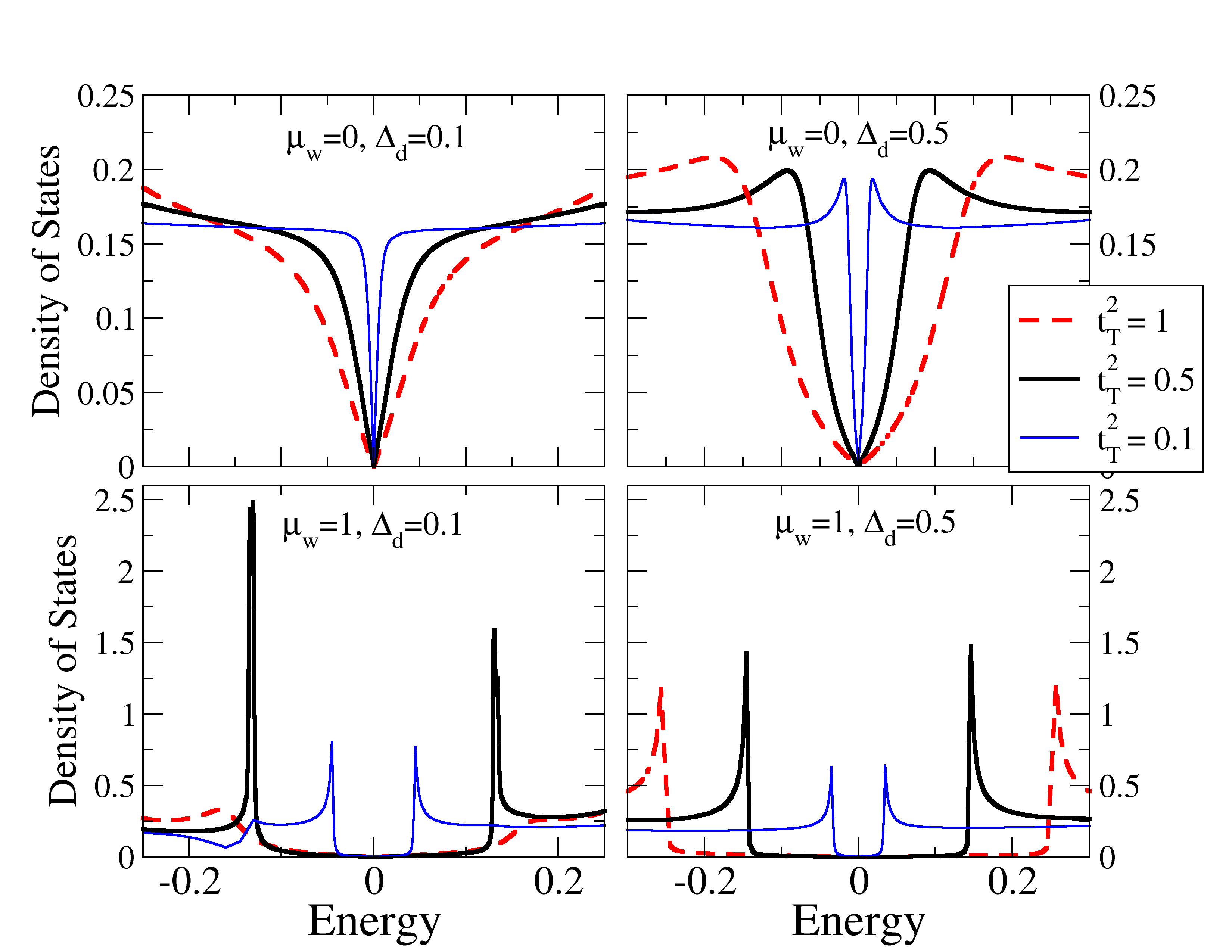}
\end{center}
\caption{\label{DOS}  Density of states of the wire at $\mu_s=0$ for different values of $\Delta_d$, and the coupling $t_T$. In the lower right panel, the density of states mimic that of a fully gapped system. However, this   picture is not correct, because subgap states states are always present (cfr. other panels), even though their number is exponentially suppressed as a function of $|t_T|^2$, as explained in the text.   }
\end{figure}

In Fig. \ref{DOS} we plot the density of states of the wire close to its Fermi energy for different values of the Hamiltonian parameters.  We are mostly interested in the induced superconducivity, thus, in this part of the paper, for the sake of simplicity, we use negligible spin-orbit coupling and focus on the density of states of the wire very close to the Fermi energy.

Conventional d-wave superconductors present coherence peaks at $\omega \sim \Delta_d$, there is no gap in the excitation spectrum, and subgap states manifest in a linear behavior 
 of the  density of states for $\omega<\Delta_d$, and it is exactly zero only at the Fermi level.
Proximity induced superconductivity in the wire, inherits all these properties, as clear from the plots in Fig.s \ref{DOS}.
In agreement with  Fig. \ref{sigma}, the stronger is the coupling $t_T$, the larger is the distance between the coherence peaks, and as a consequence, 
a more robust superconductivity is induced in the wire.

Fine tuning of the chemical potential of the wire with respect to the one of the superconductor, e.g. by gating the structure, can allow us to improve proximity effect.
Indeed, at fixed $t_T$ and $\Delta_0$, we show that for our choice of the parameters, away from half filling (at $\mu_w= 1$) the distance of the coherence peaks and the depletion of the subgap region show that a more robust superconductivity is induced in the wire with respect to $\mu_w=0$.

\subsection{Proximity effect: peculiarities of nodal superconductors}

\begin{figure}[!h]
\begin{center}
\includegraphics[width=0.5\textwidth]{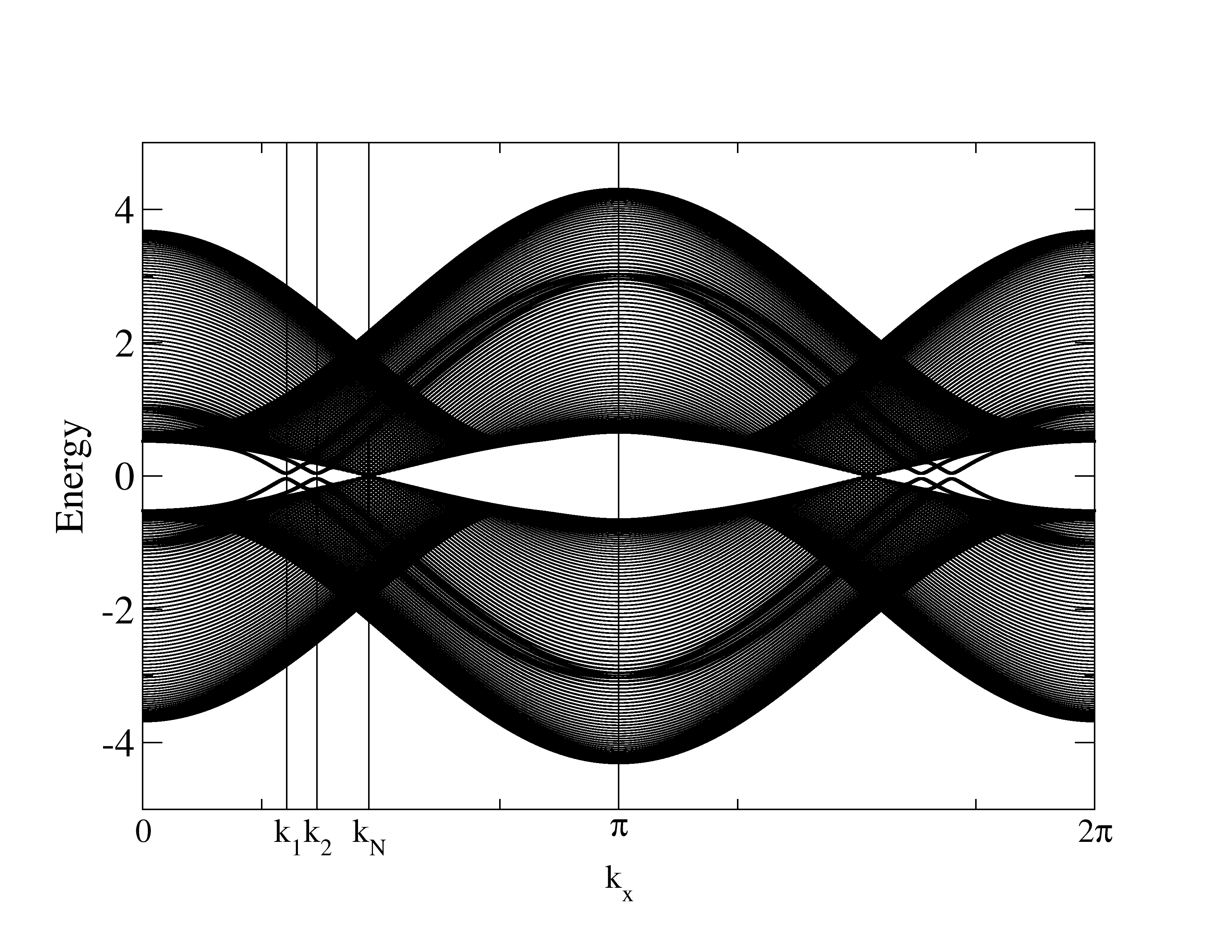}
\end{center}
\caption{\label{bands} Excitation spectrum of the full  superconductor+wire system, calculated with the self-consistent TBBDG scheme, for $\mu_{w}=-1$. The two Fermi points of the isolated wire are labeled as $k_1$ and $k_2$, in this case they both lie at the left of the superconductor nodal point $k_N$. The bands of the wire at the chemical potential, are easily recognized, they open close to the Fermi points of the isolated wire. A coupling $t_T=0.3$ and a spin-orbit interaction $\alpha=0.1$ are used in the calculation, to simulate weak coupling and strong spin-orbit interaction.  Self consistent calculations have been performed using a unit cell with $N=200$ atoms and a mesh in the Brillouin zone of $256$ k-points.}
\end{figure}

We now move to our self-consistent TBBDG calculations. 
In Fig. \ref{bands} we report a typical excitation spectrum of the wire+superconductor system, in the presence of strong Rashba spin-orbit coupling ($\alpha=0.1$), in weak coupling tunneling regime ($t_T=0.1$), 
The nodal structure of the superconductive substrate leads to a net partition of the Brillouin zone into separate regions.
This has been recognized, in the literature,  as  a key ingredient for the appearance of non trivial TRITS\cite{Zhang:2013}.
Indeed, the absence of a nodal line makes s-wave superconductors of poor relevance for the search for TRITS \cite{Haim:2016}.

Following \onlinecite{Zhang:2013}, the wire is in its topological phase if the pairing amplitude is negative in an odd number of Fermi points within $k_x \in [0, \pi ]$. Therefore it is crucial to determine the relative position of the Fermi points of the wire with respect to the superconductor nodal point. 
The nodal point $k_N$ of the superconductor, defined as the k-point in the folded 1D Brillouin zone where the gap is zero (see Fig. \ref{bands}), can be calculated as the intersection between the 2D Fermi surface $\xi^s_{k_x,k_y} = 0$ and the 2D nodal line $\Delta _{k_{x},k_{y}} = 0$. According to the present model, it is the solution of the two-equation system:
\bea
 -2t_{s} \left( \cos k_{x} + \cos k_{y} \right) &=& \mu_{s} \qquad \textrm{(Fermi surface)}\\
 k _{x} & = & k_{y} \qquad \textrm{(nodal line)}
\enea
The nodal point is thus $k_N = \arccos (-\mu_{s} /4t_{s})$. 
In the self-consistent scheme, the value of the chemical potential $\mu_s \simeq -0.3 $ corresponds to $k_N\simeq 1.49$. It is reported in Fig. \ref{bands} as a vertical line lying close to $\pi/2$. The Fermi points of the isolated wire are defined by the condition $\xi^w _{k_x}=0$, which becomes (in the presence of Rashba spin-orbit couoling):
\beq
 -2t_w \cos k_x \pm 2\alpha \sin k_x= \mu_w  \label{fermi}\:.
\eneq
The solutions of the above equation in the $[0,\pi]$ interval are reported in Fig. \ref{bands} as $k_1$ and $k_2$. A Rashba  interaction of $\alpha=0.1 $ is used, to simulate a strong spin-orbit regime, as it is the case in InAs wires  experimentally realized [\onlinecite{Bercioux:2015, Campagnano:2015, Yokoyama:2014, Lucignano:2012,Montemurro:2015, Montemurro:2015b}],  other parameters are $t_w=1$, $\mu_w=-0.3$, $t_T=0.3$. A tiny gap opens in the wire spectrum. This may seem to be not in agreement with the density of states of Fig. \ref{DOS}, clearly indicating subgap states in the wire. However, wire states are strongly hybridized with those of the underlying superconductor localized at $k \sim k_N$ down to zero energy: these states populates the subgap density of states of the wire.

\subsubsection{Band structure and pair correlation as a function of $\mu_w$}

Looking at Fig. \ref{bands}, it is clear that, if we restrict to the $k_x\in [0,\pi]$ interval, the nodal point divides the Brillouin zone into two regions, defined by  (1) $k_x<k_N$ and (2) $k_x>k_N$. In the absence of spin-orbit coupling, the wire has a single Fermi point $k_F$ lying  either in region (1) or (2). At the nodal point, the chemical potential of the wire is 
\beq
\mu_w = -2t_w \cos(k_N) = -t_w \frac{\mu_s}{2 t_s} \label{eq:muw}
\eneq
Within the present work we take $t_w=t_s=1$, therefore the Fermi point of the wire coincides with the nodal point when $\mu_w = \mu_s / 2$. In self-consistent calculations we fix $\mu_s = -0.3$, and we find that $k_F$ is located at $k_N$ when $\mu_w = -0.15$. When $\mu_w < \mu_s/2$, $k_F$ is at the left of $k_N$, otherwise $k_F$ is at the right of $k_N$.

\begin{figure}[!h]
\begin{center}
\includegraphics[width=0.45\textwidth]{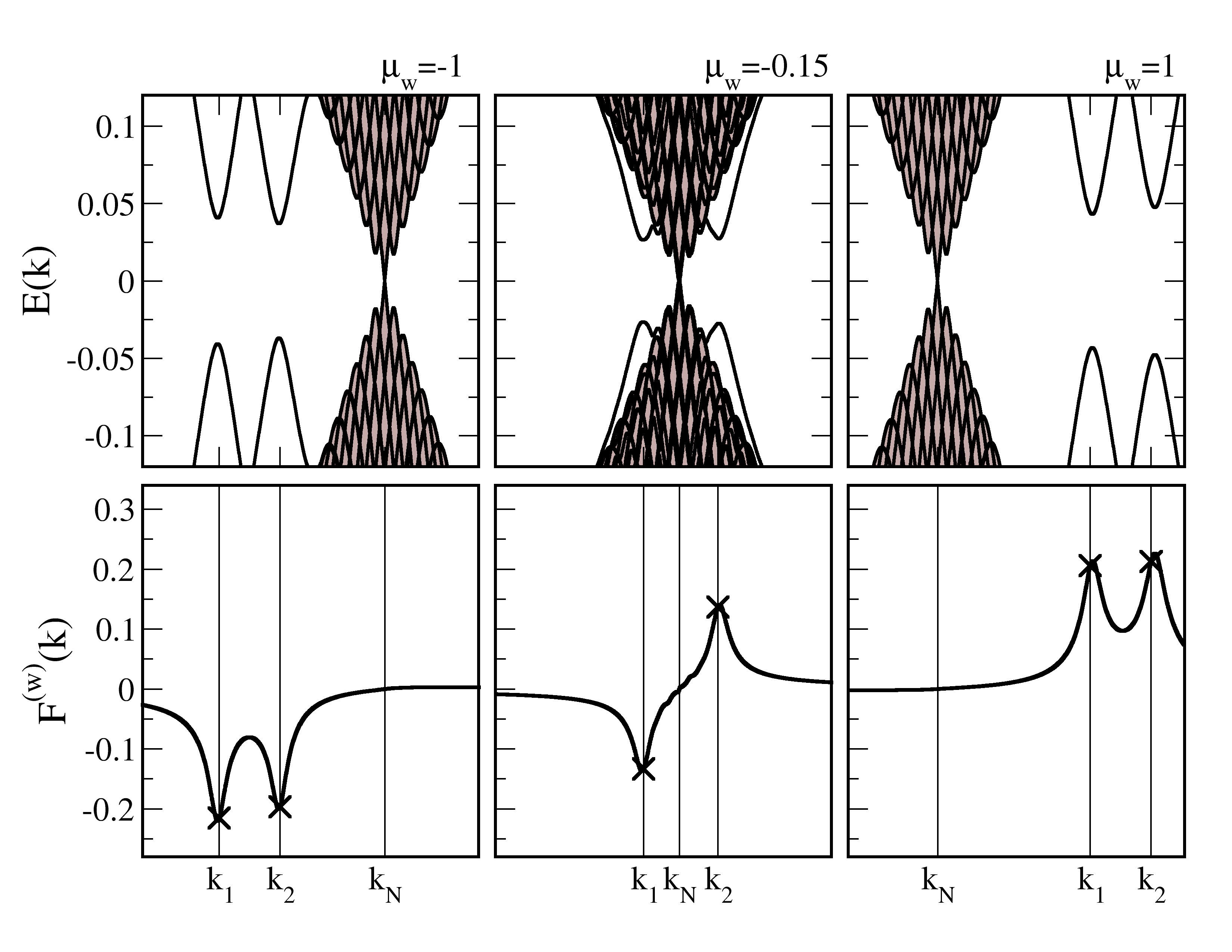}
\end{center}
\caption{\label{fpairkw}  Induced pairing reported as a function of k, upon increasing the chemical potential of the wire. Self-consistent calculations with spin-orbit coupling have been performed using the parameters  $\mu_{w} = -1,-0.15,1$, $\alpha=0.1$, $t_{T} = 0.3 $. 
Correspondingly (lower panels) we show the pairing correlation function   changing signs depending on the chemical potential. 
}
\end{figure}

The inclusion of spin-orbit coupling in the wire splits the Fermi point into two points $k_1$ and $k_2$, and their relation with $k_N$ leads to three possible scenarios: (i)$k_1<k_2<k_N$, (ii) $k_1<k_N<k_2$, (iii) $k_N<k_1<k_2$, as it is schematically shown in Fig.~\ref{fpairkw} (upper panels). 
The three panels represent a magnification of the excitation spectrum around the Fermi points. Upon increasing the chemical potential of the wire, the Fermi points change their location with respect to the superconductor nodal point, moving from the case (i) to (iii).  

In Fig. \ref{fpairkw} (upper panels) also shown is the excitation spectrum of the bulk superconductor (shadow region).  While the bulk excitations of the superconductor are poorly affected by the presence of the wire, the excitations localized on the wire feel the presence of the substrate through an opening of the gap at the Fermi points of the isolated wire.

The  pair correlation function induced on the wire, $F^{(w)}$, has been computed as a function of $k_x$, in the $[0,\pi]$ interval, for several values of $\mu_w$, and the results are illustrated in Fig. \ref{fpairkw} (lower panels). A moderate wire-superconductor coupling was considered ($t_T= 0.3 $),

The Fermi points of the isolated wire, $k_1$ and $k_2$, are indicated by crosses. They lie close to the extrema of the curve $F_{ind}(k)$, except when the Fermi points lie far from $k_N$. 
The interesting information emerging from Fig. \ref{fpairkw} is that the location of the Fermi points with respect to the superconductor nodal point fully determines the topological phase of the wire, as it is explained below.

A non trivial phase is expected when the induced pairing is negative on an odd number of Fermi points\cite{Qi:2010}. When the induced pair correlation function is positive (or negative) on both the Fermi points $k_1$ and $k_2$, the wire is in a trivial topological phase. On the other way around, an opposite sign of $F_{ind}$ at the two Fermi points leads to a non trivial topological phase. 

Figure \ref{fpairkw} shows that the sign of the pair correlation function is fully determined by the position of the Fermi points with respect to the nodal point. When $k_1$ lies in the first region ($k_1 < k_N$), the pair correlation function is negative, when $k_1$ lies in the second region ($k_1>k_N$), $F^{(w)}$ is positive. A similar discussion holds for $k_2$. Such geometrical considerations lead to a simple analytical criterion for the determination of the topological phase of the wire.
In fact, by looking at Eqs. (\ref{fermi}) and (\ref{eq:muw}), we found a non trivial TRITS if the chemical potential of the wire satisfies the following condition
\beq
 \left| \frac{\mu_w}{t_w} -\frac{\mu_s}{2 t_s}\right| \le \frac{2\alpha}{t_w} \sqrt{1-\left(\frac{\mu_s}{4t_s}\right)^2} \: .\label{eq:topo}
\eneq
In the limit $|\mu_s| \ll 4t_s$, a $4\alpha$ width for the non trivial topological region is found, in agreement with Ref. \onlinecite{Zhang:2013}, where a nodal superconductivity originating from a $s_\pm$ superconductor is hypotheisized.

\subsubsection{Band structure and pair correlation as a function of $t_T$}

\begin{figure}[!ht]
\includegraphics[width=.45\textwidth]{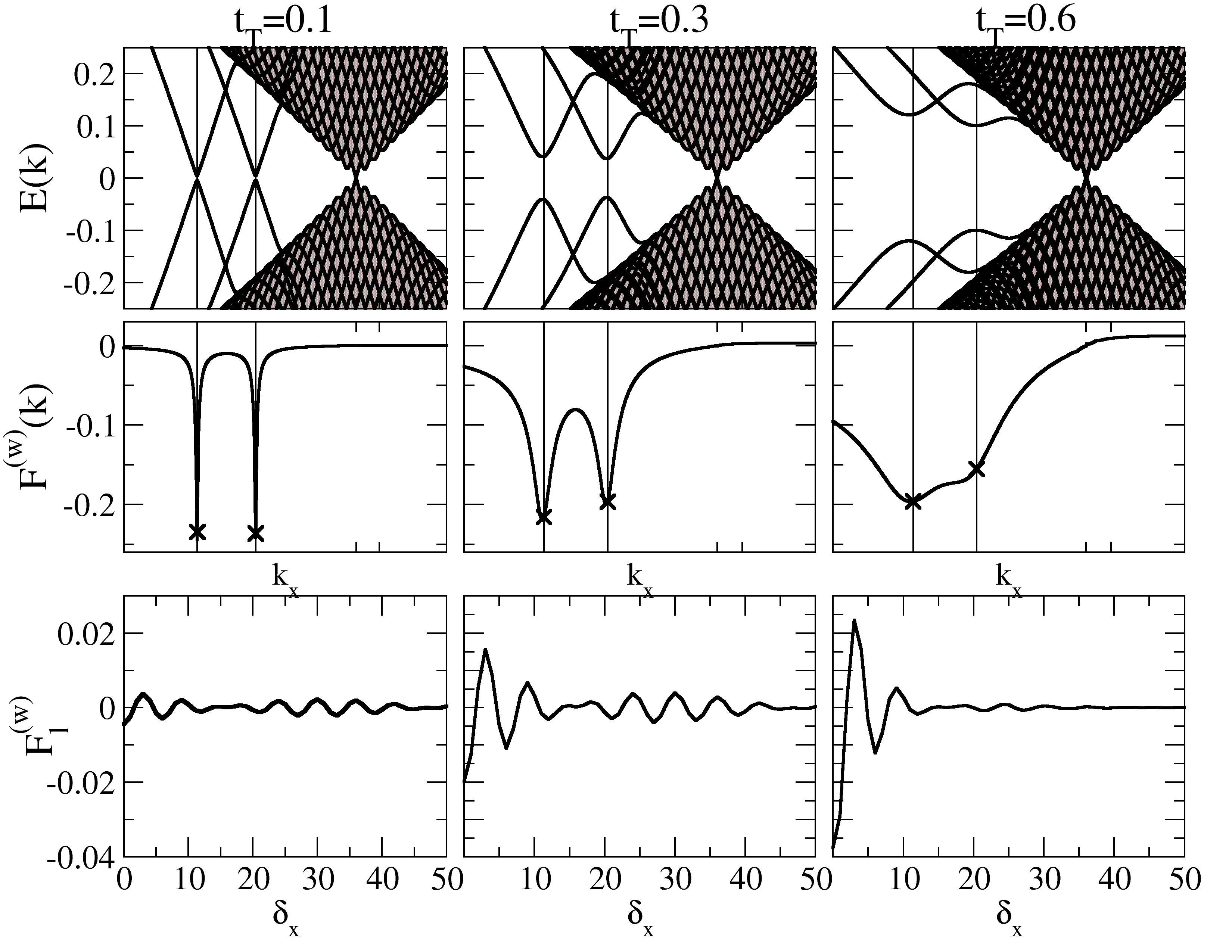}
\caption{\label{bandtt}
 Excitation spectra (upper panels), induced pair correlation function in the momentum space (middle), real space components of the pair correlation functions as a function of the distance $\delta_x$ between neighouring sites (lower panels), calculated using a self-consistent TBBDG approach, upon increasing the coupling $t_T$ between wire and superconductor ($t_T = 0.1, 0.3, 0.6$ going from left to right). The shadow region in the upper panels corresponds to the excitation spectrum of the bare superconductor. In panels are 
explicitly reported the Fermi point of the isolated wire.  The chemical potential of the wire was fixed at $\mu_w=-1$. The spin-orbit coupling is $\alpha=0.1$.}
\end{figure}

Upon increasing the coupling $t_T$ between the wire and the superconductor, there is a transition from weak to strong coupling regime. The parameter $t_T$ has a key importance, since it can be experimentally modified by a suitable engineering of the interface between the intrinsic superconductor and the wire, leading to a possible technological realization of TRITS. Thus we perform self-consistent TBBDG calculations of the band structure and the induced pair correlation function, upon increasing $t_T$. The results are illustrated in Fig. \ref{bandtt} for d-wave superconductivity.

The weak coupling regime leads to negligible effects on the position of the  Fermi points, while, in strong coupling regime the Fermi points might be significantly displaced by proximity effect.
In the middle panels of Fig. \ref{bandtt} the induced pair correlation function is reported as a function of $k_x$.
It can be noted that $F_{ind}$ has two extrema at the Fermi points of the isolated wire (cross symbols). Upon increasing the coupling, its shape changes from a Dirac-like function to a sinusoidal curve, with a significant increase of the broadening.  So, it could be misleading to try getting an estimation of the induced gap just looking at the pair correlation function at the Fermi point, since the value of $F_{ind}(k_F)$ is quite insensitive to the coupling with the substrate. More interesting, instead, is to Fourier transform the pair correlation function and look at the real space components $F^{(w)}_l$ (lower panels of Fig. \ref{bandtt}).
At very weak couplings, the correlations have a long range and the real space components are spread over many neighbors. By increasing the coupling, only a few nearest neighbor correlations survive with increased strength. Thus explicitly accounting for the proximity effect, tells us that superconducting correlations in the real space can be even long ranged. Thus, local , $F^{(w)}_0$, and extended, $F^{(w)}_1$ correlations are the most relevant but do not always give a complete description of the induced superconducting phase. 

\subsubsection{Spin-orbit coupling}

\begin{figure}[!h]
\begin{center}
\includegraphics[width=0.45\textwidth]{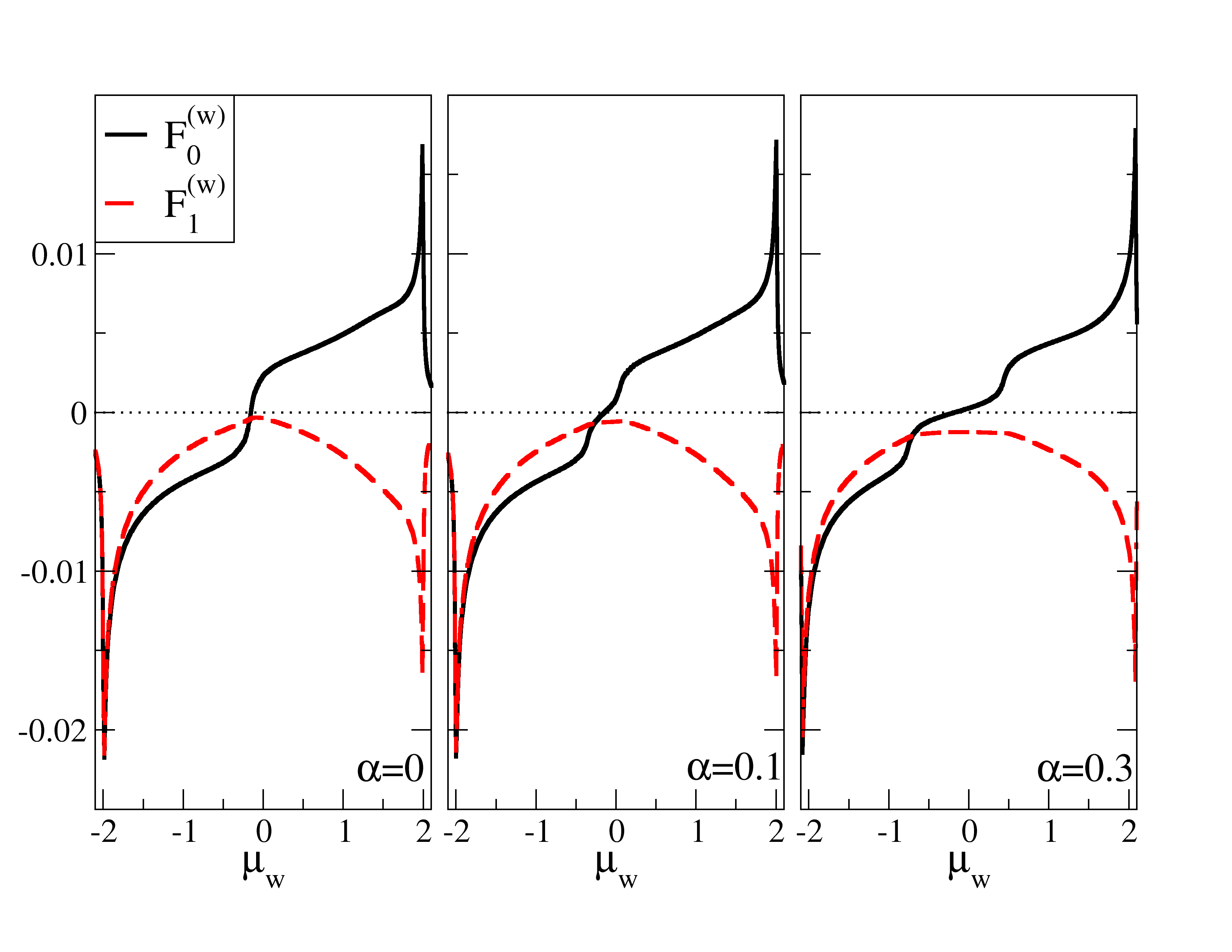}
\end{center}
\caption{\label{locs}Pair correlation function induced on the wire by a d-wave superconductor in weak coupling regime. Local and extended s-wave components are shown upon changing the chemical potential of the wire $\mu_{w}$. Self-consistent calculations have been performed using the parameters  $\mu_{s} = -0.3$, $t_{T} = 0.1$. The local component is vanishing at $\mu_{w} = \mu_{s}/2$. }
\end{figure}

Local, $F^{(w)}_0$, and extended, $F^{(w)}_1$, s-wave components of the induced pair correlation function have been calculated as a function of $\mu_{w}$, and the results are shown in Fig. \ref{locs}. The two components have the same weight at the band edges (same sign at the bottom, opposite sign at the top of the band), but the extended component is always negative, while the local component feels a change of sign at $\mu_{w}=\mu_{s}/2$. In Fig. \ref{locs} the dependence on the spin-orbit coupling strength is reported. By  increasing $\alpha$, there is an increase of the width of the central region around $\mu_{w}=\mu_{s}/2$, where the local component is smaller (in absolute value) than the extended component. 
As it is known from the literature\cite{Zhang:2013}, the extended component is responsible of the appearance of a non trivial TRI topological phase. Upon increasing the spin-orbit coupling, the extended component increases in the non trivial topological region.

\subsection{Criteria for the formation of Majorana bound states: Topological phase diagram in the weak  coupling regime}

A non trivial phase is expected  in the wire when the induced pairing is negative on an odd number of Fermi points\cite{Qi:2010}. 
Such phase, in the case of a finite wire, would display Kramers Majorana doublets at the two ends of the wire.
As reported above, in weak superconductor-wire coupling regime, the topological phase can be deduced only on the ground of geometrical considerations. If the two Fermi points of the isolated wire lie at opposite sides with respect to the superconductor nodal point, the induced pairing is negative on an odd number of Fermi points and this leads to a simple, analytical condition for the non trivial phase, given by Eq. (\ref{eq:topo}). 

\begin{figure}[!ht]
\begin{center}
\includegraphics[width=0.45\textwidth]{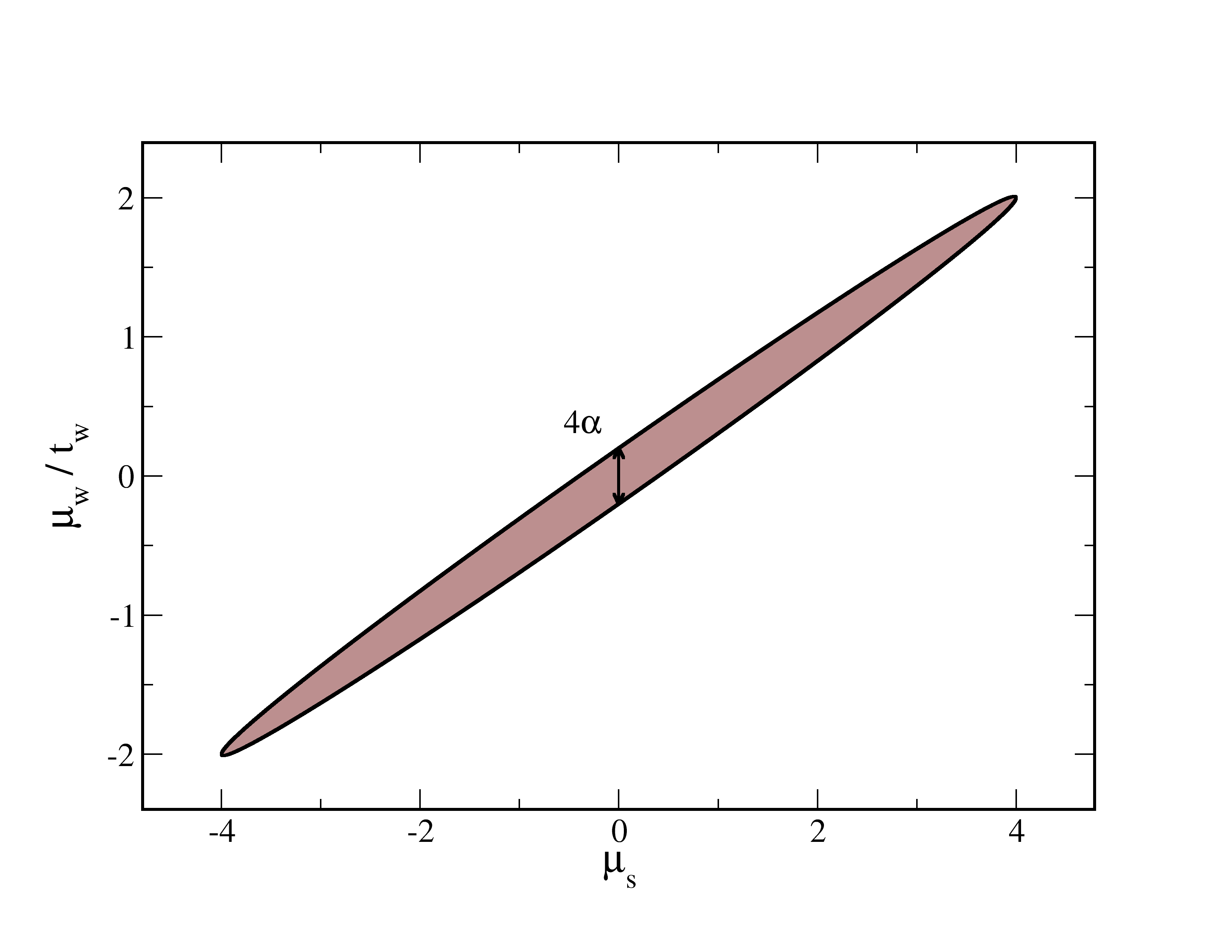}
\end{center}
\caption{\label{topo1}  Topological phase diagram in weak coupling regime, defined from Eq. (\ref{eq:topo}).  The chemical potential of the wire is plotted versus the chemical potential of the superconductor. Inside the shadow zone, the system is in a non trivial TRI topological phase. A value $\alpha=0.1$ has been used.}
\end{figure}

In Fig. \ref{topo1}, the phase diagram of the system is reported as a function of $\mu_{s}$ and $\mu_{w}$. At $\mu_{s}=0$ the criterion for the topological phase ( $-2\alpha< \mu_{w}<2\alpha$) is the same found from Zhang et al. for intrinsic, extended, s-wave superconductivity\cite{Zhang:2013}. 
The phase diagram consists of two parallel straight lines, whose distance is proportional to the spin-orbit coupling.
Upon changing the chemical potential of the superconductor, the agreement with Ref. \onlinecite{Zhang:2013} is kept in the region far from the boundaries ($\mu_s = \pm 4t_s$). Approaching the band edges (bottom and top of the band), a collapse of the topological region is observed.

The phase diagram, reported in Fig. \ref{topo1}, has been obtained analytically, according to Eq. (\ref{eq:topo}). 
In fact, our results, illustrated in previous sections, do show that the spin-orbit coupling does not modify  the shape of the excitation spectrum. The same happens for the  shape of the pair correlation function close to the Fermi points, as explained in the following section. The main effect of the spin-orbit coupling consists of a translation of the Fermi points. The same phase diagram can be constructed  using the TBBDG results, just looking at the sign of the pair correlation function at the Fermi points of the isolated wire.  The results are in  agreement with Fig. \ref{topo1}.

In the strong coupling regime, the phase diagram is not that simple to be accessed, as there is a substantial shift of the Fermi points with respect to the uncoupled systems and different approaches based on the full Green's function of the systems have to be adopted, which are far from the purposes of this paper\cite{noifuturo}. 

 \subsection{Further considerations of the pair correlation functions and topological phase diagram in the weak coupling limit}
\begin{figure}[!h]
\begin{center}
\includegraphics[width=0.45\textwidth]{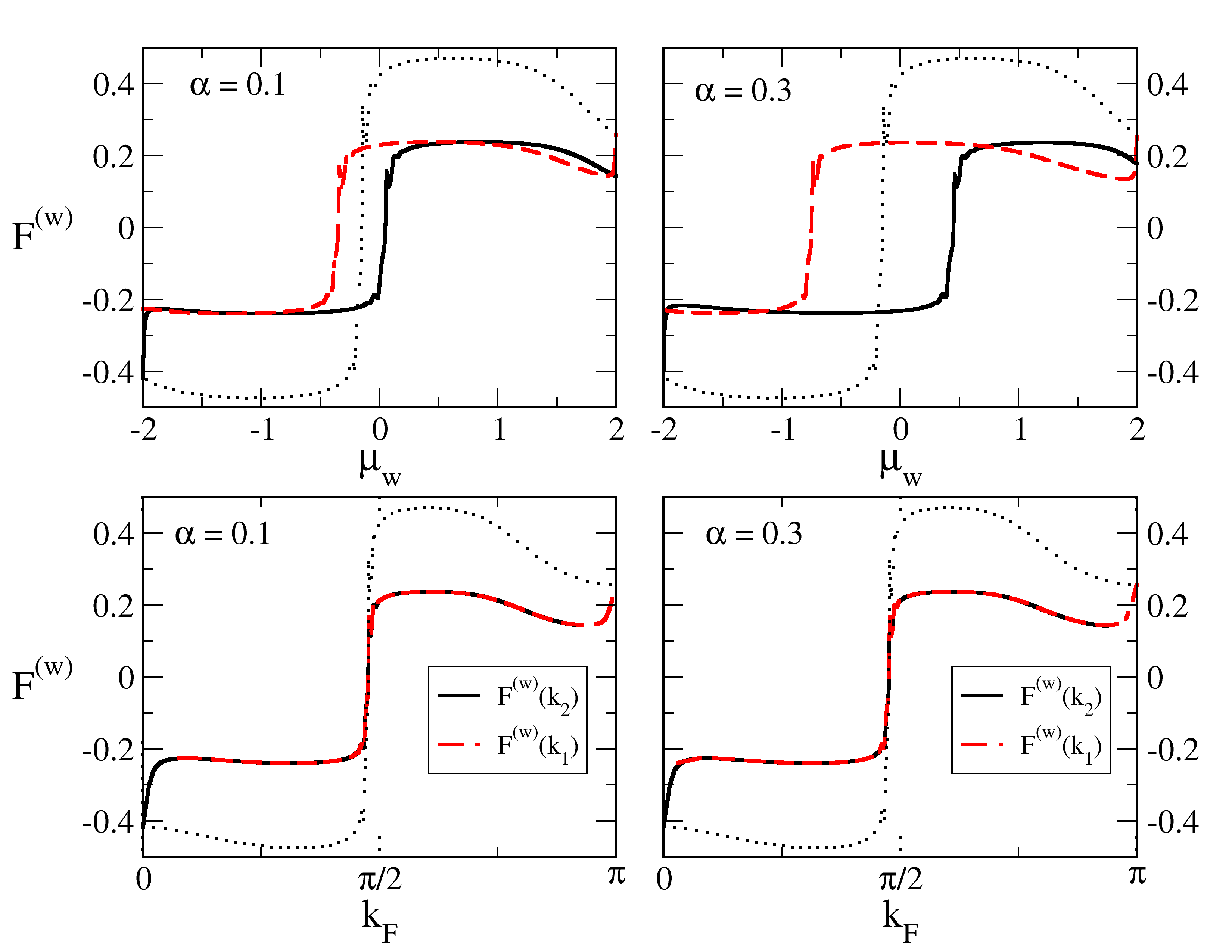}
\end{center}
\caption{\label{dind}  Induced pair correlation function, calculated at the two Fermi points of the isolated wire, plotted as a function of $\mu_{w}$ (upper panels) and as a function of $k_{F}$(lower panels), in weak coupling regime. Two different values of the spin-orbit coupling have been considered: $\alpha=0.1$ (left), $\alpha=0.3$ (right). For comparison, the results without spin-orbit coupling are also shown (dotted line). In lower panels, both the spin-orbit curves (dashed and solid lines) perfectly match. All calculations have been performed using a self-consistent scheme, with the parameters $t_s=1$, $\mu_{s} = -0.3$, $t_{T} = 0.1$.}
\end{figure}
We here compute the pair correlation function at the Fermi points of the isolated wire ($k_{1}$, $k_{2}$). 
The results are shown in Fig. \ref{dind} as a function of the chemical potential (upper panels) and as a function of the Fermi k-vector (lower panel). 
Black(solid)/red(dashed) lines correspond to the two spin-orbit bands, while the dotted line is the result without spin-orbit coupling.
The left/right panels correspond to different values of spin-orbit coupling. 
This picture summarizes all we discussed above, and several considerations are worth emphasizing. \\
1) A non trivial phase is expected when the induced pairing is negative on an odd number of Fermi points\cite{Qi:2010}. In Fig. \ref{dind} (upper panels), this criterion consists of saying that the system is in a non trivial phase when the two curves have opposite signs. And this is true  in the central region around $\mu_s/2$.  The width of this region increases upon increasing the spin-orbit coupling. The analytical expression based on geometrical arguments, defining that region is given in Eq. (\ref{eq:topo}).\\
2) The spin-orbit coupling makes the correlation weaker with respect to the results without spin-orbit coupling.
3) Once the Rashba interaction is switched on, the curves with different spin-orbit couplings are identical, apart from a rigid shift of Fermi points. This is even more evident in the lower panels, where the induced pair correlation is shown as a function of the k-vector. All the curves match perfectly one to the other, to indicate that the unique role of the spin-orbit coupling consists of a rigid shift.\\
4) $F^{(w)}  (k_{F})$ changes the sign at the superconductor nodal point $k_N$ (Fig. \ref{dind} lower panels).  This confirms what we anticipated above: if a Fermi point of the wire lies at the left of $k_N$, the corresponding  induced pair correlation is negative, otherwise it is positive. The topological criterion thus becomes a question of  finding where the Fermi points of the wire are located with respect to the superconductor nodal point.

\section{Conclusions}
The superconductivity induced on a wire by proximity effect has been analyzed in  detail, with the implementation of a self-consistent tight-binding code for the resolution of the Bogoliubov de Gennes Hamiltonian, complemented with  the calculation of self-energy and Green's function using a path-integral approach.

The TBBDG scheme allowed us to study the whole system (superconductor+wire), and to compare it either to the isolated wire or the bare substrate. We observe some interesting features emerging from our analysis. 

In the weak coupling regime the Fermi points of the wire are only poorly shifted with respect to the case of an isolated wire. Following Ref.~\onlinecite{Qi:2010} the system is in a non trivial topological phase when the induced pairing is negative on an odd number of Fermi points. Since spin-orbit coupling has the only effect to split the  Fermi points of the wire, it is easy to verify whether the criterion is satisfied, or not. This  leads us to a simple, analytic, expression determining the topological phase diagram of the system as a function of the chemical potentials of the superconductor and the wire, and the spin-orbit coupling interaction $\alpha$, as reported in Eq. \ref{eq:topo}. 
Pair correlations induced on the wire, consist of localized functions centered at the Fermi points in the momentum space. This corresponds to real-space superconducting correlations extended to many neighbors, as shown in lower panels of Fig. \ref{bandtt}. By contrast, in Ref. \onlinecite{Zhang:2013} the authors assume only local and nearest neighbor superconducting pairings. However, our topological criterion is consistent with that found in Ref. \onlinecite{Zhang:2013} when $\mu_s \ll t_s$, and extending it to a range of parameters accessible within our approach. 

At stronger coupling  the pair correlation functions get broadened in the k-space, and the wire  Fermi points get displaced with respect to the isolated wire, and eventually proliferate. This makes the study of the topological phase diagram more involved, and it will be the subject of further studies \cite{noifuturo}.

To conclude, HTS  are likely to be good candidates for TRITS, but the experimental conditions have to be carefully chosen, in order to maximize the induced superconductivity by proximity effect. 
In contrast to   s-wave superconductors,  d-wave superconductors have the required features for generating Majorana modes without the introduction of external magnetic fields, and this could open the way to new experimental realizations of  Majorana states.

\begin{acknowledgments}
We acknowledge enlightening discussions with  Y. Nazarov and F. Tafuri as well as financial support from FIRB 2012 project "HybridNanoDev" (Grant No.RBFR1236VV).
\end{acknowledgments}

\appendix
\section{Density of states}
The density of states of the wire can be calculated from the retarded 
Green's function of the wire as:
\begin{equation}
\rho^w(\omega)=-\frac{1}{2\pi}\Im m\mbox{Tr} \left(\hat{\tau}_x \hat{\mathcal{G}}_{ret}^w\right) \,.
\end{equation}
In order to calculate the wire's Green's function we need to compute the retarded self-energy. 
This can be done by substituting $\imath\omega \rightarrow \omega+\imath0^+$ in the Matsubara expression of 
Eq.(\ref{selfenergy}).
The calculation can be done using the relation
\begin{equation}
\frac{1}{f(x)+ \imath 0^+}=\mathcal{P}\frac{1}{f(x)}-\imath\pi\delta(f(x)) \,.
\label{dirac-id}
\end{equation}
\begin{widetext}
After a straightforward integration of Eq.~\ref{selfenergy} we obtain, in the case $\Delta_0=t_s=1$, for $\hat{\Sigma}_{ret}=\Re e \hat{\Sigma}_{ret} +i  \Im m \hat{\Sigma}_{ret}$
\begin{eqnarray}
\Im m \hat{\Sigma}_{ret}(k_x,\omega)=\frac{|t_T|^2}{8} \left[ \frac{\omega^2}{8}-\cos(k_x)^2\right]^{-\frac{1}{2}}
 \left[ 1-\frac{\omega^2}{8}+\cos(k_x)^2\right]^{-\frac{1}{2}}
 \mbox{sign}(\omega)
 \times\\
 \theta\left[\frac{\omega^2}{8}-\cos(k_x)^2\right] \theta\left[ 1-\frac{\omega^2}{8}+\cos(k_x)^2\right][-\omega \hat{1}+ 2 \cos(k_x)(\hat{\tau}_x+\hat{\tau}_z)]\: ,\nonumber
\end{eqnarray}
and
\begin{eqnarray}
\Re e \hat{\Sigma}_{ret}(k_x,\omega)=-\frac{|t_T|^2}{8}\left\{ \left[\frac{\omega^2}{8}-\cos(k_x)^2\right] 
\left[\frac{\omega^2}{8}-\cos(k_x)^2-1\right]\right\}^{-\frac{1}{2}} 
\times\\ \left\{ \theta\left[\frac{\omega^2}{8}-\cos(k_x)^2\right]- \theta\left[ 1-\frac{\omega^2}{8}+\cos(k_x)^2\right]\right\}
[-\omega \hat{1}+ 2 \cos(k_x)(\hat{\tau}_x+\hat{\tau}_z)]\: .\nonumber
\end{eqnarray}
Similarly, one can compute the self-energy for an arbitrary value of $\Delta_0$. The calculation is slightly  
more involved but nevertheless the result can be presented in a closed form.
Let us introduce the following functions:
\[
v_{1,2}=\frac{\cos(k_x) (\Delta_0^2-1)\pm\sqrt{\Delta_0^2[\omega^2/4-4\cos^2(k_x)]+\omega^2/4}}{1+\Delta_0^2} \, .
\]
When $\omega^2 < 16 \Delta_0^2 \cos(k_x)^2/(1+\Delta_0^2)$ the 
roots $v_1$ and $v_2$ are complex and the self-energy is given by

\begin{equation}
 \hat{\Sigma}_{ret}(k_x,\omega)=\frac{|t_T|^2}{4(v_2-v_1)(1+\Delta_0^2)} 
\left\{\frac{1}{\sqrt{1-1/v_1^2}}\left[\frac{\hat{\alpha}}{v_1}+\hat{\beta}\right]-
 \frac{1}{\sqrt{1-1/v_2^2}}\left[\frac{\hat{\alpha}}{v_2}+\hat{\beta}\right] \right\} \, ,
\end{equation}
where the matrices
\[
\hat{\alpha}=-\omega \hat{1}+2\cos(k_x)( \hat{\tau}_z+ \Delta_0\hat{\tau}_x) \: ,
\]
and 
\[
\hat{\beta}=2\hat{\tau}_z-2 \Delta_0\hat{\tau}_x \,,
\]
have been introduced.
For  $\omega^2\ge16 \Delta_0^2 \cos(k_x)^2/(1+\Delta_0^2)$, using again the relation in Eq.(\ref{dirac-id}) we obtain for the real part of the self energy:
\begin{equation}
\Re e \hat{\Sigma}_{ret}(k_x,\omega)=\frac{|t_T|^2}{4(v_1-v_2)(1+\Delta_0^2)}  \left\{\frac{\hat{\alpha}+v_1\hat{\beta}}{\sqrt{v_1^2-1}}\left[ \theta (-1-v_1)-\theta (v_1-1) \right] \right.
 \left.-\frac{\hat{\alpha}+v_2\hat{\beta}}{\sqrt{v_2^2-1}}\left[ \theta (-1-v_2)-\theta (v_2-1) \right] \right\} \, .
\end{equation}
For the imaginary part of the self-energy we have 
\begin{equation}
\Im m \hat{\Sigma}_{ret}(k_x,\omega)=|t_T|^2 \frac{\mbox{sign}(\omega)}{4|v_1-v_2|(1+\Delta_0^2)} \left\{\frac{\hat{\alpha}+v_1\hat{\beta}}{\sqrt{1-v_1^2}} \, \theta (1-v_1^2) 
+   \frac{\hat{\alpha}+v_2\hat{\beta}}{\sqrt{1-v_2^2}} \theta (1-v_2^2) \right\} \,.
\end{equation}
\end{widetext}



\end{document}